\documentclass[conference]{IEEEtran}
\usepackage{graphicx}

\graphicspath{ {./images/} }
\usepackage{float}
\usepackage{xcolor}
\usepackage{listings}
\usepackage{svg}
\usepackage{tikz}
\usetikzlibrary{decorations.pathreplacing,shapes,trees,positioning,chains,fit,calc}
\usepackage{pgfplots}
\usepackage{pgfplotstable}
\definecolor{greennmap}{RGB}{0, 153, 25}

\definecolor{DarkGreen}{HTML}{008000}
\definecolor{DarkBlue}{HTML}{000080}
\definecolor{DarkMagenta}{HTML}{660E7A}
\lstdefinestyle{PythonStyle}{
	language=Python,
	basicstyle=\ttfamily\small,
	columns=fullflexible,
	showstringspaces=false,
	keywordstyle=\color{DarkBlue},
	stringstyle=\color{DarkGreen},
	morekeywords={yield}
}

\begin{document}
\title{A Framework for Threats Analysis \\Using Software-Defined Networking}
\author{
        \IEEEauthorblockN{Francisc Moldovan\IEEEauthorrefmark{1}\IEEEauthorrefmark{2}, Ciprian Opri\c{s}a \IEEEauthorrefmark{1}\IEEEauthorrefmark{2}}
        \IEEEauthorblockA{\IEEEauthorrefmark{1}Bitdefender\\
        \IEEEauthorrefmark{2}Technical University of Cluj-Napoca
        \\\{fmoldovan, coprisa\}@bitdefender.com}        
}

\IEEEoverridecommandlockouts
\IEEEpubid{\makebox[\columnwidth]{978-1-5386-8445-0/18/\$31.00~
\copyright2018 IEEE \hfill} \hspace{\columnsep}\makebox[\columnwidth]{ }}

\maketitle

\begin{abstract}
The ability to analyze network threats is very important in security research. Traditional approaches, involving sandboxing technology are limited to simulating a single host, missing local network attacks. This issue is addressed by designing a threat analysis framework that uses software-defined networking for simulating arbitrary networks. The presented system offers flexibility, allowing a security researcher to define a virtual network that is able to capture malicious actions and to be restored to the initial state afterwards. Both the framework design and common usage scenarios are described. By providing this framework, we aim to ease the analysis effort in combating cyberthreats.
\end{abstract}

\section{Introduction}
Most cyberthreats nowadays propagate through computer networks, bypassing the borders of a single host. Modern threat intelligence must consider this aspect while analyzing new threats or new malware. Observing the behavior of a given threat, isolated on a single host is not enough, as some malicious actions are only performed in a real-world network.

Traditional malware research uses virtual machines for dynamic analysis \cite{willems2007toward}. Virtual machines are handy because they represent a controlled environment, where suspicious software can be run and the system changes can be observed, while having the ability to restore the environment to the original state.

For instance, we can analyze a ransomware sample \cite{kharraz2015cutting,oprisa2013ransomware} by running it in a virtual machine. The ransomware will perform some malicious actions, like encrypting files and leaving ransom notes. The analysis (automated or manual) can detect these actions, either by monitoring the ransomware actions or by comparing the state of the machine after the infection with the initial state. After the analysis have been performed, the original snapshot can be restored, preparing the VM for another analysis. In case of a backdoor or bot \cite{rieck2008learning}, besides the actions performed on the infected host, a complete analysis should also examine the network interactions, as this type of malware communicates with a Command and Control server located on the Internet. 

However, there are malware samples and hacking tools that are not only directed towards the Internet but also look inside the local network for potential targets. If the simulated environment is limited to a single host, connected or not to the Internet, little information can be gathered regarding the network behavior of the analyzed threat. The current paper proposes a framework designed to overcome this limitation, by extending the sandbox machine concept to a sandbox network.

We achieve the proposed goal by employing software-defined networking \cite{benzekki2016software}, a concept that allows defining a virtual network programmatically, offering both elasticity and scalability.

The next section presents similar approaches that we came across while designing the framework presented in this paper. Section~\ref{sect:arch} presents the framework architecture, both network-wise and functionality-wise. The experimental results follow, where we present some common network attacks successfully detected by our framework and some considerations regarding performance. The paper ends with the conclusions section.

Among our contributions, worth mentioning are the fact that virtual networks created using the framework reside on the system itself with high level of interaction with host OS firewall, routes and interfaces, while still preserving the much entailed level of isolation of the host from potentially malicious virtual network activity. Also, easy configuration of hosts prior to network creation proves facile by using high level abstraction at the created framework level. 

\section{Related Work}	

The \textit{mininet} SDN Framework\cite{team2017mininet}, appeared in 2012, and as its authors concisely affirm: 
\textit{"Mininet creates virtual networks using process-based virtualization and network namespaces - features that are available in recent Linux kernels. In Mininet, hosts are emulated as bash processes running in a network namespace, so any code that would normally run on a Linux server (like a web server or client program) should run just fine within a Mininet "Host"."}. It became quite popular and mature, hence already being available for usage as both a standalone, custom built virtual machine, and as a software package available on official Linux repositories, such as the \textit{apt}, used by Debian-based distributions. As the official page of \textit{mininet} shows, a considerable number of publications surfaced in the domain of SDN thanks to their implementation. 

By comparing and contrasting our approach with \cite{team2017mininet}, a series of obvious differences emerge. While \cite{team2017mininet} provides a vast range of functionalities, our emphasis on security research using SDN fails to benefit greatly from the framework. First of all, bypassing network isolation proves a difficult task, and the user needs to rely solely on the \textit{APIs} available in the framework itself. While we also assured isolation of a virtual network, our solution also focuses on comprehensive interaction between a virtual network and the underlying host operating system. More exactly, our focus lead to the following: enabling easy addition to the network of virtual machines or real ones, toggling Internet connectivity of the network, easy reconfiguration of network structure without an impact on performance. An analysis on scalability with \textit{mininet} has been carried out in \cite{mininet}, proving the efficiency of using \textit{lightweight virtualization} for creating virtual networks, an encouraging aspect in validation of our implementation.

The authors of \cite{sdnsecu} place emphasis on utility of SDN in security research, and also shed light upon the fact that popularity of SDN increases faster among networking professionals than with security researchers. Their publication brings strong arguments backed up by existing examples from the SDN security community, showing the advantages of using virtual networks in securing networks in a novel manner. Among the SDN systems that bring novelty to security, worth mentioning are dynamic control of malicious or suspicious network flows, a centralized monitoring system for detection of network flooding or network anomalies, and even the development of network programming languages for easy deployment. While our approach utilizes SDN for security research, \cite{sdnsecu} presents use of SDN for securing a real network, aspect which shall prove useful for future work. 

An interesting approach towards network security using SDN is presented in \cite{deception}. The authors make use of the popular virtual network  framework \textit{mininet}\cite{team2017mininet}
in order to perform \textit{reconnaissance deception}. Basically, by leveraging the power of software defined networking, each real host in a network presents to the others a bogus image of the network, once taking part in a network scanning activity. While network deception fails to replace security scanning, the mechanism greatly aids in increasing the time needed for an insider to infer the layout and structure of the real network. Compared to our solution, \cite{deception} shows an interesting usage of an existing framework, while we illustrate the creation of a framework for creating virtual networks for use in security research.

\section{Framework Architecture}
\label{sect:arch}
\subsection{Overview}
While designing and experimenting with the SDN framework, a series of compulsory requirements have emerged. The first aspect, and probably the most critical: it should 
be able to run on commodity hardware, including a personal laptop, while in the same time having the ability to model 
hundreds of real network hosts, each with customizable services exposed (e.g. FTP, HTTP, SSH).

An issue that surfaced early on in the testing phase is that emulation of services fails
to deliver realistic results, since many exploits only utilize their
malicious payloads on detection of real, very specific services. Basically, emulated services represent bogus output exposed to attackers, which perfectly matches the one of a real application. During reconnaissance of their target network and systems, most attackers resort to \textit{banner grabbing}. This technique, as described in \cite{kondo2014penetration}, consists of connecting to a remote application using \textit{network Swiss army knife} tools such as \textit{nc}, described in \cite{skoudis2006counter}. In most cases, exact version of software surfaces to the malicious scanner. Usually, service emulation solutions also cover banner grabbing, together with response to common commands specific to a network service. Despite these efforts, many automatic bots rarely get tricked into triggering their malicious act.

The SDN framework provides elasticity in the usage for creating a virtual network, and for the scenarios tested, we chose 3 virtual machines, namely: \textit{victim}, \textit{attacker}, \textit{scanner}. The number of network hosts, though, can easily be extended to the hundreds, if not thousands without using additional VMs. Moreover, the additional hosts are able to expose real services on the network, with the particularity that they must be installed on the \textit{victim} VM. 

Making use of not only simple, non-VM virtual hosts, but also VMs in our framework stems from service emulation negatively affecting expected detection outcome. In order to obtain scenarios as realistic as possible, these hosts need to reveal real network services, with the possibility of vastly extending their variety. As a \textit{victim} machine hosting various network services, we chose the popular \textit{Metasploitable}, which as its creators affirm, represents \textit{"a test environment [...] a secure place to perform penetration testing and security research"}. While virtual machines deliver excellent feedback in security research, using more than 3 VMs on a typical laptop can seriously take a toll on the host machine functionality.

The other two virtual machines in our test setup, namely \textit{attacker} and \textit{scanner}, were also chosen with regard to their specificity. The \textit{attacker} VM represents an installation of an x86 Windows OS. The motivation behind our choice lies in the fact that many malware families exist for Windows, and the scope of our framework targets exactly having the ability to collect network traffic information with malware or malicious network activities. In addition, we have installed some popular pentesting tools compiled for Windows, such as network scanning tool \textit{nmap} and  brute force password cracking tool  \textit{Hydra}\cite{allen2012advanced}. The first one aids in performing discovery and analysis of neighboring hosts and their services in a network, while the second one employs automation by leveraging the power of dictionary attacks. Finally, the \textit{scanner} VM represents a version of the well-known Debian-based Linux distribution, \textit{Security Onion}\cite{burks2012security}. It comes as a great aid in detection of malicious network activities, by using a series of specific, pre-installed IDS tools such as Bro, Suricata or Snort. In addition, the authors of \textit{Security Onion} encourage the security researcher utilizing this distribution to spend days, weeks tailoring the stack of dedicated security tools, for example by adding new signatures for detection.

Worth mentioning in the end are the fact that the \textit{attacker} virtual machine could be used in order to model the attacker from inside the virtual network, while \textit{victims} are the \textit{victim} virtual machine together with all the non-VM, \textit{virtual redirector nodes}, that map certain ports onto ports of services that \textit{need to be installed on the victim machine}.

\subsection{Network-wise Architecture}
At the core of the SDN Framework lie features in the Linux Kernel itself. First of all, the core of a virtual network represents the Linux Bridge, which acts as a gateway for the virtual network itself. With security in mind, both of the host computer, and of a real network bearing the system as one of its hosts, we carefully chose Linux Firewall rules and chains, in order to have isolation of the virtual network. On the other hand, the virtual network may be configured to have Internet access, but DNS configuration at host level remains a task for the human actor, in order to avoid accidental Internet leaks.

\begin{figure}[h]
	\includegraphics[width=\columnwidth]{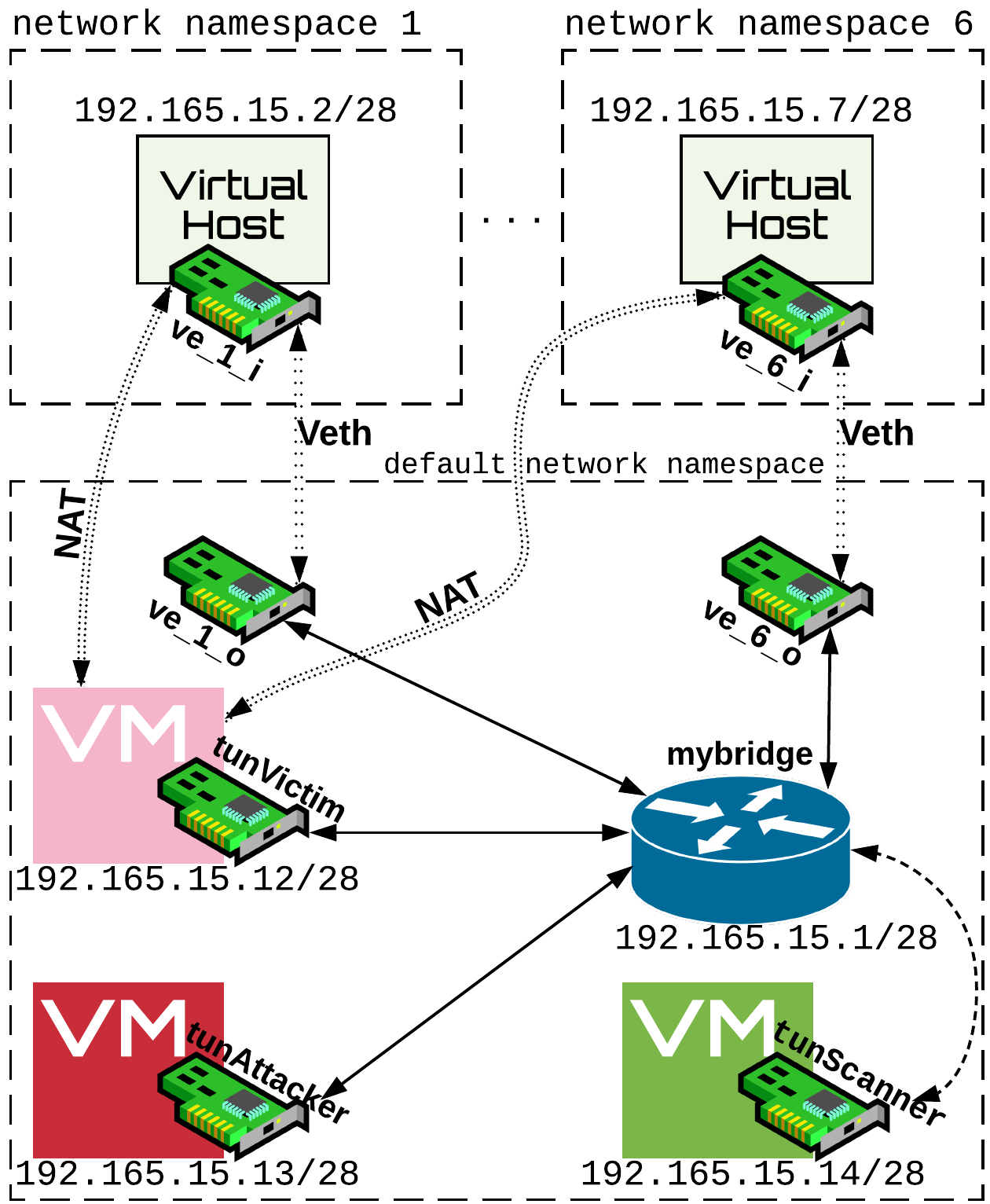}
	\caption{An overview of a virtual network implemented using our
		framework. The \textit{VM}-label rectangles represent the following, by color: \textit{pink-victim}, \textit{red-attacker},
		\textit{green-scanner}. Next to each NIC symbol is the interface name in Linux. \textit{mybridge} represents the virtual network bridge.}
	\label{fig:fig1}
\end{figure}

Linux Network Namespaces and Linux virtual Ethernet devices lie at the heart of each non-VM host in the network. Concerning namespaces in the Linux operating system, the \textit{Linux Manpages}\cite{man} entry for \textit{namespaces} provides relevant pieces of information: \textit{"A namespace wraps a global system resource in an abstraction that
makes it appear to the processes within the namespace that they have
their own isolated instance of the global resource.  Changes to the
global resource are visible to other processes that are members of
the namespace, but are invisible to other processes.  One use of
namespaces is to implement containers"}. 
Basically, a virtual, non-VM host in our SDN solution represents a \textit{network namespace} having a network interface that connects using virtual Ethernet to a \textit{veth pair} interface in the default network namespace, which in turn connects to the virtual network bridge.

Since each non-VM host in the virtual network we created represents solely a networking context, 
we benefit from \textit{lightweight virtualization}, allowing the existence of as many such hosts as the network address range allows us, without there appearing noticeable impact on performance. Furthermore, leveraging the capabilities of \textit{NAT} (\textit{Network Address Translation}), each such host allows a mapping of its own port to a port of the \textit{victim} machine, in order to deceive the attacker that the virtual host actually exposes certain services on the network on certain ports, while in reality they exist on the \textit{victim} virtual machine, this being a requirement. 

With automation in mind to an extend as vast as possible, we chose to utilize networking capabilities of the underlying host operating system. Therefore, the 3 virtual machines, \textit{victim}, \textit{attacker} and \textit{scanner} programmatically receive DHCP lease reservation upon creation or modification of the virtual network. 

A particularity of the chosen setup aims at not only obtaining an isolated network, but also assuring in-depth scanning of the network traffic. In order to do so, the so-called \textit{scanner} virtual machine connects to the virtual network in \textit{promiscuous mode} and also benefits from \textit{port mirroring} such that all the network traffic from the bridge gets replicated in the \textit{scanner} interface. A significant aspect that entails attention represents \textit{hiding} the \textit{scanner} machine, while at the same time assuring that it can communicate with the \textit{bridge}, in order to receive a copy of its entire traffic, and also the DHCP configuration. The solution for this represents \textit{blocking} all \textit{ARP} packets, except those coming from the bridge, at the level of the \textit{scanner} VM.

Worth mentioning is the method via which the components mentioned above connect with one another. Each virtual machine receives a \textit{tun/tap interface} onto which it attaches. Next, this interface connects to the \textit{bridge} of the virtual network. 
At the level of non-VM hosts, namely the network namespaces which contain solely a network interface configuration and NAT redirection rules, communication takes place using so-called \textit{veth pairs}. More exactly, upon creation, a network namespace receives a network interface which communicates via a tunnel with a pair interface in the default network namespace. In turn, the \textit{veth pair interface} connects to the virtual network bridge (as seen in Fig. \ref{fig:fig1}).

\begin{figure}[h]
	\includegraphics[width=\columnwidth]{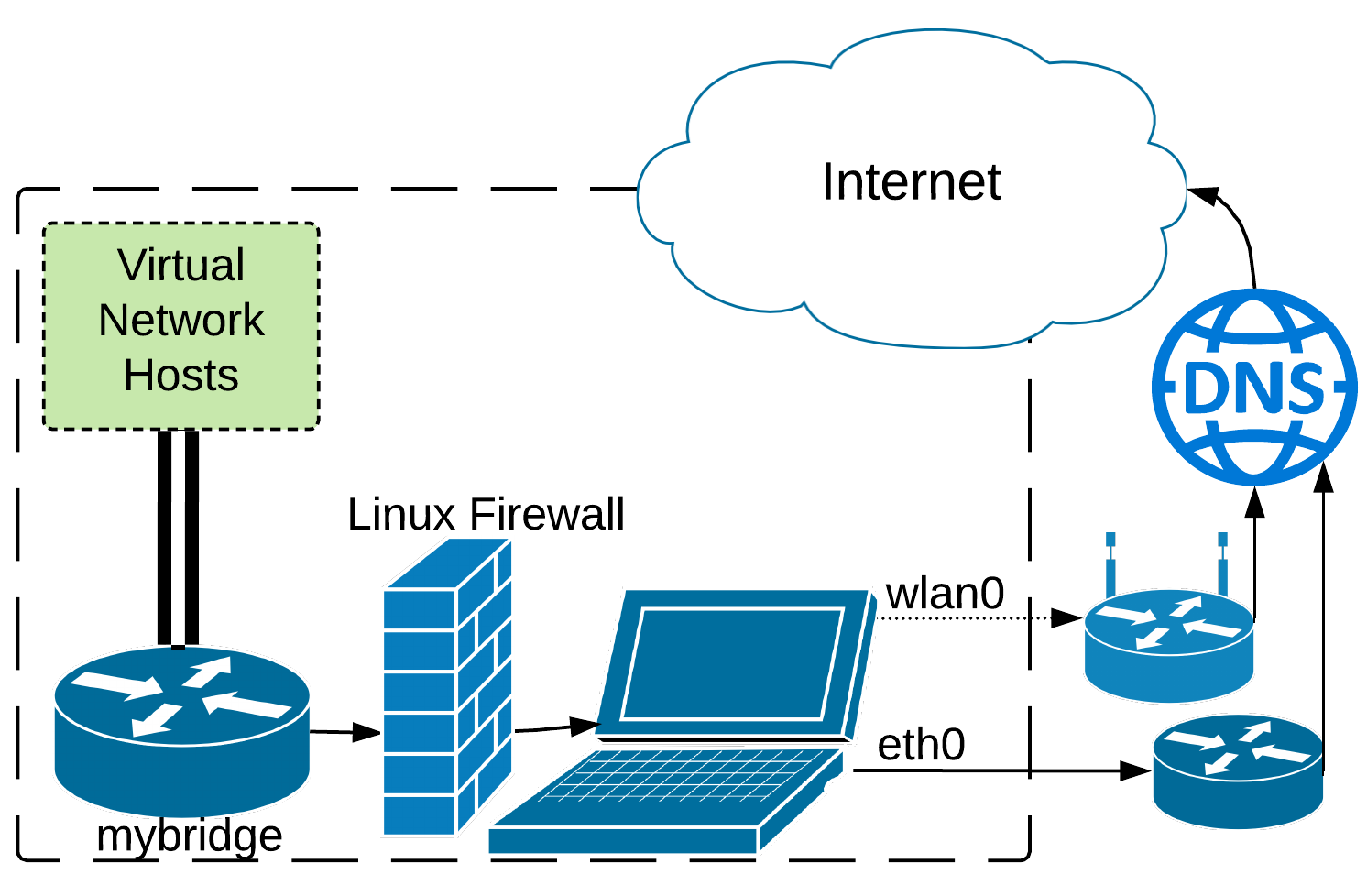}
	\caption{An illustration of the interfaces and the firewall of the host laptop running Linux. The devices in the area delimited by the interrupted rectangle lie on the OS of the laptop itself.} 
	\label{fig:fig2}
\end{figure}

As far as networking is concerned, the critical aspect of Internet access requires debate. While most SDN solutions, such as \textit{Mininet}\cite{mininet} place network 
isolation high on their list of priorities, the security researcher
utilizing our framework might make it adamant that their setup requires connections towards outside networks. The entire architecture from Figure \ref{fig:fig1} lies encapsulated in a laptop running Linux. Namely, upon creation of the virtual network, the laptop presents the following interfaces: \textit{mybridge}: the virtual network gateway, \textit{3 tun/tap interfaces}: for the 3 virtual machines in our test setup, and a number of \textit{veth} interfaces equal to twice the number of virtual, non-VM hosts. For obvious reasons, on a complete virtual network setup, only the veth interfaces from the default namespace appear when listing the interfaces on the laptop (Figure \ref{fig:fig1} shows exactly the encapsulations at network namespace level). The other half of veth interfaces became configured accordingly to the virtual network and pushed in their specific network namespaces corresponding to each non-VM host. Given the external network access issue described earlier, and also the structure of the underlying system, let us commence with describing the flow of outbound connections originating in our virtual network.

Based on Figure \ref{fig:fig2}, we can observe that the Linux Firewall lying on the host operating system establishes whether the virtual network we created can initiate communication towards an external network via Ethernet or WiFi. Note that complete isolation may exist, and that in our approach, having Internet connectivity works only with outbound connections, all inbound traffic being blocked.

\subsection{Functionality-wise Architecture}
For better illustrating the \textit{functionality-wise architecture}, an attack scenario and the following steps shall be described. The security researcher runs an infected file on the \textit{attacker} virtual machine. The sample ran 
aims at scanning the local network. It shall detect as victims the \textit{victim virtual machine} and the non-VM, \textit{virtual redirector nodes} that solely map onto their own ports services that must be installed on the \textit{victim} machine. The \textit{scanner} virtual machine receives the entire network traffic and could also trigger detections. The security researcher performs a complete reset of the virtual machines, but not before collecting the network capture, which needs manual inspection for new attack signatures to be added to the \textit{scanner virtual machine}.

Functionality-wise, from its inception phase, our framework aimed at ease of use, automation, and possibility of reconfiguration even when the virtual machines already performed a complete boot-up, without the requirement of their restart.

One of the advantages our solution brings, represents the series of validations prior to creation of the virtual network. More precisely, MAC addresses of all the interfaces to be created and added must not match any existing ones on the system. At IP level, validation takes place such that there exists no overlap of the IPv4 range of a network-to-be and existing ones. 

An Object-Oriented abstraction was added, such that end users can utilize the framework with disregard to specificities underlying services invoke. Implementation-wise, we chose to utilize Python, due to ease in performing network-specific calculations, such as inferring the number of hosts, gateway IPv4 address or performing programmatic inquiry of existing network interfaces on the host machine.

Since our solution utilizes a series of \textit{virtual machines}, we chose as virtualization software the Linux build of VirtualBox, as it provides the much entailed Command-Line Interface interaction. Moreover, it being open-source assures
ease of use backed up by a large on-line community.

Programmatically, the user needs to provide a series of scenario-specific parameters, such as IPv4 address of the network and its netmask, MAC addresses, names of network interfaces and also NAT redirections. For our series of scenarios, the user also needs to provide configuration parameters for the collection of NICs in Figure \ref{fig:fig1}, except the NICs responsible for interconnecting the network namespaces to the default network namespace, since they receive unique automated names and MAC addresses. By default, a virtual network created using our framework prepares the network and the VMs, there existing the requirement that the user chooses which of the available IPv4 addresses to allocate to virtual, non-VM hosts, and also what service redirections to perform at the level of each one (earlier we described that each one can provide an entire collection of redirections towards real services installed on the \textit{victim} VM). Taking into account that the network aims at being used by a human actor in security research, DHCP reservations represent a feature. In our series of runs, we chose a convention that \textit{victim}, \textit{attacker} and 
\textit{scanner} receive the last 3 IP addresses in the network to be created. 

Provided parameters choice took place, current state of each VM gets assessed, after which follows their network-wise configuration at OSI layer 1 and 2, regardless of them being switched on or off. In our specific case, \textit{tun/tap} adapters receive the desired identification MAC and name. For our runs, we then prompted the user with the task of choosing the \textit{visibility} of the virtual network. More exactly, as Figure \ref{fig:fig2} shows, one must choose a NIC of the host device for assuring connectivity towards the Internet. At this step there exists the possibility of choosing complete isolation of the virtual network, taking into account the possibility of malicious traffic occurring during its use afterwards. Next follows the configuration of firewall and interfaces, after which the software calculates IPv4 specific parameters, and prepares a list of available IPv4 addresses.

Having presented the architecture of our framework both network-wise, and at software level, we continue by illustrating functionality of the API exposed.
As stated earlier, given the variety of use cases, there exist no restrictions 
regarding the structure of the virtual network to be created. Let us consider a scenario of the security researcher creating a network containing the 3 virtual machines mentioned earlier. Initially, the network of choice for creation might be characterized by the input parameters presented in Figure \ref{fig:userConfig}.

\begin{figure}[h]
\begin{lstlisting}[style=PythonStyle]
bridgeName = "mybridge"
tunTapNameVictim = "tunVictim"
tunTapNameAttacker = "tunAttacker"
tunTapNameScanner = "tunScanner"
testIP                = "192.165.15.0"
testNM                = "255.255.255.240"
testMACBridge         = "00:50:56:c0:aa:01"
testMACTapVictim      = "00:60:67:34:12:44"
testMACTapAttacker    = "00:60:67:34:12:55"
testMACTapScanner     = "00:60:67:34:12:66"
\end{lstlisting}
\caption{User added configuration parameters for the virtual network to be created.}
\label{fig:userConfig}
\end{figure}

The framework proceeds with configuration of all virtual hosts, followed by employing the user-established level of isolation of the network to be created. With a successful run of the solution, the output ought to be similar to the one illustrated in Figure \ref{fig:infoInferred}.

\begin{figure}[h]
\small
\begin{verbatim}
Mynet: 
bridgeName       : mybridge
tunTapName#1     : tunVictim 
tunTapName#2     : tunAttacker 
tunTapName#3     : tunScanner
hostInterfaceName: wlp2s0
nwAddress        : 192.165.15.0
netmask          : 255.255.255.240
_cidr            : 28
_broadcast       : 192.165.15.15
_gateway         : 192.165.15.1
_maxHosts        : 14
\end{verbatim}
\caption{Information inferred by the framework based on the input parameters given.}
\label{fig:infoInferred}
\end{figure}

The framework user needs to consider the number of additional non-VM hosts the network currently supports. In our specific use case, 11 slots left for IPv4 addresses assure a relatively limited sized network. An example of populating the network programmatically can be seen in Figure \ref{fig:exAPI}.

\begin{figure}[h]
\begin{lstlisting}[style=PythonStyle]
futureHostIP = myNet.availableIPs[0]
addVethHost(myNet, futureHostIP)
redirectPort(aNetwork = myNet, 
	     fromIP = str(futureHostIP), 
	     fromPort = "2121",
	     toIP = str(IPReservationVictim), 
	     toPort = "21")
\end{lstlisting}
\caption{An example of API usage: adding a new non-VM host, which also maps its port 2121 to port 21 of the victim machine, using NAT.}
\label{fig:exAPI}
\end{figure}

As seen below, adding hosts to a virtual network using our framework becomes accessible via custom wrapper functions which perform specific Linux networking
tasks such as creation of networking artifacts, address assigning, and performing NAT redirections towards the victim from a non-VM host.
Let us consider that a certain number of hosts were added, each exposing a number of service redirections towards the victim. If the user considers more hosts are to be added, changing the netmask from \textit{255.255.255.240} to one allowing more hosts existing, such as \textit{255.255.255.0} can take place by changing the \textit{netmask} of the network and rerunning the creation script. Hosts and their redirections coded in so far remain available on network reset.
Advantages the implementation presents are the fact that it takes a short time for reconfiguring the entire system forming our virtual network. In addition, changing the network address to a completely different one, like for example \textit{10.10.10.0} and renaming interfaces can also take place. The key for speed lies in the fact that the virtual machines need not restart for complete reconfiguration, and virtual non-VM hosts have a low footprint.

\section{Experimental Results}

\subsection{Attack Scenarios}
Our validation phase, with accent placed on threats analysis, consisted of 
creating a virtual network whose network traffic to contain specific attack types, such as \textit{brute-force}, but also reconnaissance activity, such as \textit{port scanning}.

\subsubsection{Port Redirections Towards Victim for Attacker Luring}

The listing shown in Figure \ref{fig:nmap}, taken from the \textit{attacker} machine, represents an output of a network scan, performed using the \textit{nmap} tool, with the following mapping scripted into a Python dictionary prior to virtual network creation. For each tuple seen in Figure \ref{fig:portRedirections}, the first item represents a port of a virtual redirector host, and the second one, the port of the \textit{victim} virtual machine, which needs to host the services.

\begin{figure}[h]
\begin{lstlisting}[style=PythonStyle]
redirDict[2]=[(11,21),(22,22),(33,23)]
redirDict[3]=[(44,23),(55,25),(66,53), 
                   (77,80), (88,111)]
redirDict[4]=[(99,139),(100,445),(110,514),
		   (120,514),(130, 1524)]
redirDict[5]=[(140,2049),(150,2121),(160,3306),
                   (170,3632),(180,5432)]
\end{lstlisting}
\caption{Port redirections towards victim. Each array offset represents the ordinal of a non-VM virtual network host.}
\label{fig:portRedirections}
\end{figure}

\begin{figure}[h]
\begin{lstlisting}[basicstyle=\small]
Nmap scan report for <@\textbf{192.165.15.2}@>
Host is up (0.00059s latency).
<@\textbf{\underline{Not shown:}}@> 167 closed ports
<@\textbf{\textcolor{blue}{PORT   STATE SERVICE}}@>
<@\textbf{\textcolor{greennmap}{11/tcp open  systat}}@>
<@\textbf{\textcolor{greennmap}{22/tcp open  ssh}}@>
<@\textbf{\textcolor{greennmap}{33/tcp open  dsp}}@>
<@\textbf{\underline{MAC Address:}}@> CA:99:1D:6E:E3:7D (Unknown)

Nmap scan report for <@\textbf{192.165.15.3}@>
Host is up (0.00049s latency).
<@\textbf{\underline{Not shown:}}@> 165 closed ports
<@\textbf{\textcolor{blue}{PORT   STATE SERVICE}}@>
<@\textbf{\textcolor{greennmap}{44/tcp open  mpm-flags}}@>
<@\textbf{\textcolor{greennmap}{55/tcp open  isi-gl}}@>
<@\textbf{\textcolor{greennmap}{66/tcp open  sqlnet}}@>
<@\textbf{\textcolor{greennmap}{77/tcp open  priv-rje}}@>
<@\textbf{\textcolor{greennmap}{88/tcp open  kerberos-sec}}@>
<@\textbf{\underline{MAC Address:}}@> 5E:25:B9:73:F7:B9 (Unknown)

Nmap scan report for <@\textbf{192.165.15.4}@>
Host is up (0.00049s latency).
<@\textbf{\underline{Not shown:}}@> 165 closed ports
<@\textbf{\textcolor{blue}{PORT   STATE SERVICE}}@>
<@\textbf{\textcolor{greennmap}{99/tcp  open  metagram}}@>
<@\textbf{\textcolor{greennmap}{100/tcp open  newacct}}@>
<@\textbf{\textcolor{greennmap}{110/tcp open  pop3}}@>
<@\textbf{\textcolor{greennmap}{120/tcp open  cfdptkt}}@>
<@\textbf{\textcolor{greennmap}{130/tcp open  cisco-fna}}@>
<@\textbf{\underline{MAC Address:}}@> 16:CC:9F:FA:CA:A7 (Unknown)

Nmap scan report for <@\textbf{192.165.15.5}@>
Host is up (0.00042s latency).
<@\textbf{\underline{Not shown:}}@> 165 closed ports
<@\textbf{\textcolor{blue}{PORT   STATE SERVICE}}@>
<@\textbf{\textcolor{greennmap}{140/tcp open  emfis-data}}@>
<@\textbf{\textcolor{greennmap}{150/tcp open  sql-net}}@>
<@\textbf{\textcolor{greennmap}{160/tcp open  sgmp-traps}}@>
<@\textbf{\textcolor{greennmap}{170/tcp open  print-srv}}@>
<@\textbf{\textcolor{greennmap}{180/tcp open  ris}}@>
<@\textbf{\underline{MAC Address:}}@> 32:6F:E3:2C:47:86 (Unknown)
\end{lstlisting}
\caption{An example of port scanning output with a virtual network we built with custom NAT bindings to real services existing on the \textit{victim} virtual machine.}
\label{fig:nmap}
\end{figure}

\subsubsection{Brute Force and Port Scanning Detection}
A test scenario for validating the functionality of our framework involved creating a number of 4 virtual hosts that perform NAT redirection on port 21 towards the \textit{victim} virtual machine. From the \textit{attacker} VM, we ran a script which launches into execution 4 instances of the \textit{Hydra} brute force tool, targets being the four virtual hosts. Results shown on the security panel of the \textit{scanner} VM appear in a timely manner, showing the detection: \textit{ET SCAN Potential FTP Brute-Force attempt} with the source IP being a each virtual host, and the \textit{victim} VM as destination.

Port scanning also triggers an alarm in our \textit{scanner}, promiscuous-mode connected VM, and what is more, on launching a complex scan from the \textit{attacker} VM (see \cite{allen2012advanced}), many attack signatures trigger reactions at the level of the \textit{scanner} VM, including detection of binary payloads.

\subsection{Performance}
The first scenario for performance evaluation consists of analyzing the time required to add increasingly more virtual non-VM hosts to our network.
We chose the network address \textit{192.165.0.0/16}, and 
added a number of \textit{998 virtual hosts}, each performing the following 5 NAT service redirections:

\begin{itemize}
	\item 99/tcp  filtered metagram
	\item 100/tcp filtered newacct
	\item 110/tcp filtered pop3
	\item 120/tcp filtered cfdptkt
	\item 130/tcp filtered cisco-fna
\end{itemize}

The resulting number of hosts in the virtual network, counting the bridge, the three VMs and out non-VM virtual hosts, is 1002. With IP reservation for VMs \textit{victim}, \textit{attacker} and \textit{scanner} being \textit{192.165.255.252 - 192.165.255.254}, the IP range of non-VM hosts is \textit{192.165.0.2 - 192.165.3.231}

The plot below shows the behavior in time of the host operating system, as more non-VM virtual hosts are added to the virtual network.

\begin{figure}[h]
	\centering
    	\begin{tikzpicture}
    		\begin{axis}[
    			xlabel={Number of hosts},
    			ylabel={Time (s)},
    			grid=major
    		]
    			\addplot+[mark size=1pt] table{tables/performance.dat};
    		\end{axis}
    	\end{tikzpicture}
    	\caption{Running time by the number of hosts added to the network.}
    	\label{fig:performance}
\end{figure}
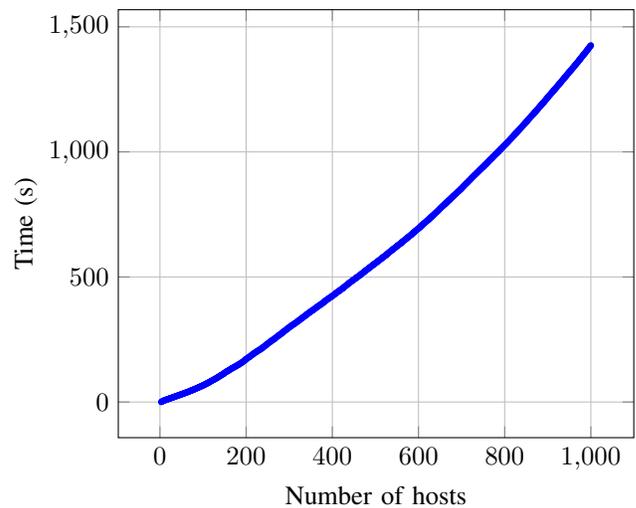

While the plot from Figure \ref{fig:performance} shows that adding a virtual redirector non-VM host takes roughly one second, this is because the host operating system requires a considerable number of transactions to take place while configuring routes, interfaces and rules. In addition, resetting the network state only targets virtual machines, since virtual redirecting non-VM hosts solely redirect to services installed onto the \textit{victim} virtual machine. The similar framework Mininet \cite{mininet} requires a similar amount of time for virtual non-VM host creation.

The second scenario consists of automated running of malware Windows PE infected samples, in order to obtain a capture of their network traffic. The time required to run a sample depends on the activity of the infected program. With the virtual network already created, once malware activity took place, it only takes a few seconds to collect the network traffic capture file, and restore state of virtual machines to their pre-attack status, using snapshots. Therefore, within an hour, leaving each malware sample a time frame of 5 minutes, and with a time of 5 seconds to restore virtual machine snapshots for \textit{victim}, \textit{attacker} and \textit{scanner} and save a *.pcap network capture from \textit{scanner} or \textit{bridge} level, an hour would allow collecting network traffic from 11 infected programs. While appearing as a tedious process, malware research usually needs to allocate time for malware to manifest, 5 minutes being considered a reasonable timespan.

\section{Conclusions}
Sandboxes are powerful tools for dynamic cyberthreats analysis but analyzing a threat in an isolated virtual machine is not enough to reveal the entire attack potential. We have proposed a framework that addresses this issue by employing software-defined networking concepts.

Our framework can be used to define a custom virtual network that resembles an entire home network or even a corporate network. The framework architecture was presented both network-wise and functionality-wise, describing the configurations that a user can perform in order to achieve the desired results.

Experimental results showed that the proposed system is able to detect common network attacks, while being able to simulate even large networks (1000 virtual hosts) on commodity hardware.

\section*{Acknowledgment}
Research supported, in part, by EC H2020 SMESEC GA \#740787 and EC H2020 CIPSEC GA \#700378.


\bibliographystyle{IEEEtran}
\bibliography{ftausdn}

\end{document}